\documentclass[aps,prl,preprint,groupedaddress]{revtex4}
\usepackage{graphicx}
\usepackage{amsmath}

\begin{document}


\title{Near-IR Subwavelength Microdisk Lasers}

\author{Q. Song$^1$, H. Cao$^1$, S. T. Ho$^2$, G. S. Solomon$^3$}
\affiliation{$^1$ Department of Applied Physics, Yale University, New Haven, CT 06520-8482 \\
$^2$ Department of Electrical Engineering and Computing Science, Northwestern University, Evanston, IL, 60208 \\ 
$^3$ Atomic Physics Division, NIST, Gaithersburg, MD 20899-8423. }

\date{\today}

\begin{abstract}

We report single-mode lasing in subwavelength GaAs disks under optical pumping. The disks are fabricated by standard photolithography and two steps of wet chemical etching. The simple fabrication method can produce submicron disks with good circularity, smooth boundary and vertical sidewalls. The smallest lasing disks have a diameter of 627 nm and thickness of 265 nm. The ratio of the disk diameter to the vacuum lasing wavelength is about 0.7. Our numerical simulations confirm that the lasing modes are whispering-gallery modes with the azimuthal number as small as 4 and a modal volume of $0.97\left(\lambda/n\right)^{3}$. 

\end{abstract}


\maketitle


Subwavelength laser sources have important applications to nanophotonic circuits, on-chip optical interconnects, very local chemical and biological sensing. Semiconductor microlaser resonators based on photonic crystals or Bragg mirrors can attain subwavelength modal volumes, but require the overall sizes of feedback structures at least several times larger than the wavelength. \cite{vahala03,painter99,akahane03} Metallic coatings provide stronger confinement of light and consequently higher device-packing density. \cite{hill07,manolatou08,mizrahi08} Surface plasmon waveguides have been exploited for terahertz quantum cascade lasers with all three dimensions smaller than the vacuum emission wavelength. \cite{fasching07, chass07, dunbar07} The dissipative loss of metal at optical and near-IR frequencies, however, is much higher than that at terahertz. Lasing at near-IR is realized in three-dimensional (3D) subwavelength metal-coated cavities at low temperature where the metal absorption is reduced. \cite {hill07} Even at 10 K the cavity quality ($Q$) factor is only a couple of hundreds, limited by the dissipative loss of metal. Another promising candidate for subwavelength laser resonators is the dielectric odisk. It has small cavity size and high $Q$ factor, because light is confined by total internal reflection at the disk edge. \cite{mccall92} To avoid high optical bend losses in small disks, most microdisk lasers have diameter over 1 $\mu$m. \cite{levi93,baba97} Recently Zhang {\it et al} have realized visible submicron disk lasers which operate at room temperature. \cite {zhang07} The smallest disks they obtain lasing have the diameter of 645 nm, which is equal to the lasing wavelength in vacuum. In addition to the cavity size, the fabrication cost is also an important factor for practical applications. The submicron-scale lasers are fabricated by electron-beam lithography and dry etching. Although it gives good control of cavity shape and surface roughness, such fabrication method makes mass production nearly impossible.  

In this letter, we report lasing in subwavelength GaAs disks at near-IR frequency. The disks are fabricated by standard photolithography and two steps of wet chemical etching. The submicron disks have good circularity, smooth boundary and vertical sidewall. Single mode lasing is obtained by optical pumping. The gain is provided by the wetting layers of  InAs quantum dots (QDs) embedded in the GaAs disks. The diameter of the smallest lasing disks is 627 nm, which is about 30\% smaller than the vacuum lasing wavelength. The 3D finite-difference time-domain (FDTD) calculations show that the lasing modes in the subwavelength disks are whispering-gallery modes (WGMs) with the azimuthal number $m$ = 4 or 5. To our knowledge, the GaAs disk laser with diameter 627 nm and thickness 265 nm is the smallest dielectric disk laser that has been reported so far.  Our simple fabrication method provides the opportunity for high-throughput production of the ultrasmall lasers. 

The sample is grown on a GaAs substrate by molecular beam epitaxy. The layer structure consists of 1000 nm Al$_{0.68}$Ga$_{0.32}$As and 265 nm GaAs. Inside the GaAs layer there are six monolayers of InAs QDs equally spaced by 25 nm GaAs barriers. Standard photolithography is used to define circular patterns of diameter 3 - 5 $\mu$m within the photoresist (PR). Then GaAs and AlGaAs are etched nonselectively in a mixture of HBr : H$_{2}$O$_{2}$ : H$_{2}$O with the ratio 4 : 1 : 25. The etching is isotropic, leading to an undercut of GaAs beneath the PR mask. In the case of a large etch depth, the diameter of GaAs disk is significantly smaller than that of the PR disk. By increasing the etch time, we reduce the GaAs disk diameter to submicron.  To obtain a fine control of the etch depth, we slow down the etching process by placing the etch solution in an ice bath. For a typical etch depth of 2.4 $\mu$m, the etch time is increased from 1 minute at room temperature to 28 minutes at 5$^{\circ}$ C. Finally, dilute HF is used to selectively etch Al$_{0.7}$Ga$_{0.3}$As, forming a pedestal underneath the GaAs disk. 
 
Figure 1(a) is a top-view scanning electron microscope (SEM) image of a GaAs disk. The disk diameter $d$ = 627 nm, much smaller than that of the PR mask ($\sim$ 5 $\mu$m). Despite of a large undercut of GaAs, the circular disk shape is well kept.  The tilt-view SEM image in Fig. 1(b) shows that the sidewall of GaAs disk is vertical and smooth. As compared to dry etching, wet chemical etching creates less surface damage. The top diameter of Al$_{0.7}$Ga$_{0.3}$As pedestal is about 260 nm. The dimension of the pedestal is set to minimize light leakage to the GaAs substrate while maintaining acceptable heat sink for the GaAs disk. The surface of pedestal in fig.1b is rough possibly due to As passivation in the selective etching process.    

In the lasing experiment, the microdisks are optically pumped by a mode-locked Ti:Sapphire laser (pulse width = 200 fs, center wavelength = 800 nm, pulse repetition rate = 76 MHz). The sample is cooled to 10 K in a liquid Helium cryostat. A long-working-distance objective lens focuses the pump beam to a single disk. Emission from the disk is collected by the same objective lens and directed to a half-meter spectrometer with a liquid nitrogen cooled CCD array detector. Figure 2 shows the measurement results of a GaAs disk with $d$ = 722 nm and thickness 265 nm.  At low pumping level, radiative recombination of carriers in the InAs QDs and the wetting layers produce spontaneous emission in the wavelength ranges of 890-950nm and 860-880nm respectively. Figure 2(a) shows the spectra of emission from the wetting layers when the incident pump power $P$ is varied from 39 to 91 $\mu$W. For $P < 50 \mu$W, the spectrum exhibits only broad-band spontaneous emission. Once $P$ exceeds 50 $\mu$W, a narrow peak appears at wavelength $\lambda$ = 865 nm. Its amplitude grows rapidly when the pump power is above 70 $\mu$W. Figure 2(b) is a logarithmic plot of the spectrally-integrated intensity $I$ of the peak versus $P$. The curve exhibits a $S$-shape with two kinks at $P \simeq$ 70 $\mu$W and 160 $\mu$W. Below the first kink, the slope of $\log I$ over $\log P$ is approximately 0.91, indicating a linear increase of peak intensity with pumping. Above the first kink, the slope changes to 3.6. The super-linear increase of  $I$ with $P$ result from optical amplification. Above the second kink the slope changes back to 0.92 due to gain saturation. Such behavior confirms lasing action in this disk which has all three dimensions smaller than the lasing wavelength in vacuum. The incident pump power at the lasing threshold is about 70 $\mu$W. The inset of Fig. 2(a) is a plot of the linewidth $\Delta \lambda$ of lasing peak versus $P$. $\Delta \lambda$ decreases drastically as $P$ approaches the lasing threshold. At threshold $\Delta \lambda \simeq 0.7$ nm. Above threshold $\Delta \lambda$ starts increasing.  This increase is attributed mostly to temporal shift of lasing frequency. \cite{Wegener95} The short pump pulses generate hot carriers in the GaAs barriers which subsequently relax to the InAs wetting layers and quantum dots. The carrier distribution keeps changing in time during the short lasing period following the pump pulse. \cite{LuoAPL01} Consequently the refractive index changes in time, causing a continuous redshift of lasing frequency. \cite{Jahnke95} In our time-integrated measurement of lasing spectrum, the transient frequency shift leads to a broadening of lasing line. At higher pumping level, the spectral shift and broadening of lasing line is more dramatic.


To identify the lasing mode, we calculate numerically the resonant modes in the subwavelength disk. Since its radius is only slightly larger than its thickness, the disk can no longer be considered as a two-dimensional cavity with an effective index of refraction. Instead we solve the Maxwell's equations in three-dimension with the FDTD method. The sizes and shapes of GaAs disks and Al$_{0.7}$Ga$_{0.3}$As pedestals are obtained from the top-view and tilt-view SEM images. Good circularity of the disk shape allows us to use the rotational symmetry to simplify the calculation. \cite{chen96} Experimentally the disks are uniformly pumped, thus the lasing modes correspond to high-$Q$ modes of frequencies within the gain spectrum. We calculate the resonant modes with good quality in the wavelength range of InAs wetting layer emission. In particular, we focus on the modes predominately TE polarized, which are similar to the lasing modes observed experimentally. For the disk of diameter 722 nm and thickness 265 nm, a whispering gallery mode with the azimuthal number $m$ = 5 and radial number $l$ = 1 is found at wavelength $\lambda \approx 860$ nm. It is very close to the wavelength of the observed lasing mode. The calculated $Q$ factor of this mode is about 4000. The surface roughness and slight shape deformation of the disk are not taken into account in the calculation, which would  degrade the $Q$ factor. Since the mode is predominately TE polarized, i.e. the magnetic field component normal to the disk plane ($H_z$) is much larger than the in-plane component, the calculated spatial distribution of $H_z$ in the cross-section of disk and pedestal is shown in Fig. 3 (a). It is evident that the mode is concentrated near the GaAs disk edge and has little overlap with the Al$_{0.7}$Ga$_{0.3}$As pedestal. Our simulations confirm that a fine control of the pedestal size is crucial to minimize light leakage through the pedestal to the substrate as well as the scattering loss caused by the pedestal surface roughness. The modal volume is calculated to be approximately $1.3\left(\lambda/n\right)^{3}$. Because of the small cavity size, the frequency spacing of high-$Q$ modes well exceeds the gain bandwidth. To achieve lasing, it is important to fine tune the disk diameter so that one of the high-$Q$ mode overlaps with the gain spectrum. 

We also calculate the $Q$ values for WGMs with $l = 1$ but different $m$ in the subwavelength disk. As shown in Fig. 3(b), the $Q$ factor drops quickly from over 10,000 to 100 when $m$ decreases from 6 to 3. The $Q$-spoiling is attributed to a dramatic increasing of light leakage through tunneling at the disk boundary. For $m = 4$, $Q \simeq 700$. To check experimentally whether we can achieve lasing in the WGM with $m = 4$, we fabricate a GaAs disk of diameter 627 nm and thickness 265 nm (shown in Fig. 1). The WGM with $m = 4$ in this disk coincides with the gain spectrum of InAs wetting layer. We observe the onset of lasing action at the incident pump power $P$ = 220 $\mu$W. The inset of Fig. 4(a) shows the lasing peak at $\lambda = 870$ nm. A threshold behavior is clearly seen in the growth of peak intensity with pumping [Fig. 4(a)]. Figure 4(b) is a plot of the peak width $\Delta \lambda$ versus $P$. $\Delta \lambda$ first decreases with increasing $P$, then saturates at higher $P$. The minimal linewidth is about 1.7 nm. Our numerical simulation reveals Q factor of the WGM with $m= 4$ is $\sim$700, and the modal volume is $0.97\left(\lambda/n\right)^{3}$. The higher lasing threshold and broader linewidth are attributed to the smaller size and lower quality of this disk as compared to the previous disk with $d$ = 722 nm.  

In conclusion, we demonstrate single-mode lasing in submicron-scale GaAs disks in the wavelength range of  860 - 880 nm. The disks, fabricated by standard photolithography and two steps of wet chemical etching, have good circularity and smooth and vertical sidewalls. Both the disk diameter $d$ and thickness are smaller than the lasing wavelength $\lambda$ in vacuum. The smallest lasing disks have the ratio $d / \lambda \simeq 0.7$. The lasing modes are whispering-gallery modes with the azimuthal number $m = 4$. The modal volume is $0.97 \left(\lambda/n\right)^{3}$. Such ultrasmall lasers have potential applications in high-density photonic circuits and microscale chemical or biological sensing systems. The simple fabrication method facilitates mass production of the subwavelength disk lasers.


This work is supported partly by NIST under the Grant No. 70NANB6H6162 and by NSF under the Grant No. ECCS-0823345. This work was performed in part at the Cornell NanoScale Facility, a member of the National Nanotechnology Infrastructure Network, which is supported by the National Science Foundation (Grant ECS-0335765)


\pagebreak

{\bf Figure Captions}

Fig. 1:  Top-view (a) and tilt-view (b) scanning electron microscope images of a GaAs disk on top of a AlGaAs pedestal. The disk diameter is 627 nm, and the disk thickness is 265 nm. The green circle is a fit of the disk shape. 

Fig.2: (a) Spectra of InAs wetting layer emission from a GaAs disk of diameter 722nm. From bottom to top, the incident pump power $P$ = 39, 53, 60, 70, 78.5 , 91.5 $\mu$ W. The spectra were vertically shifted for a clear view. Inset: linewdith of lasing peak versus pump power. (b) Log-log plot of the spectrally-integrated intensity $I$ of the narrow peak at $\lambda = 860$ nm as a function of pump power $P$. The straight lines are fitted curves.
 
Fig. 3: (a) Calculated spatial distribution of magnetic field component normal to the disk plane $H_z$ of a WGM in the cross section of GaAs disk and Al$_{0.7}$Ga$_{0.3}$As pedestal. The GaAs disk has a diameter of 722 nm and thickness of 265 nm. The radial number of the WGM $l = 1$, and the azimuthal number $m = 5$. (b) Calculated $Q$ factors of the WGMs with $l = 1$ and $m = 3, 4, 5, 6$. 

Fig. 4: (a) Linear plot of the lasing peak intensity vs. the incident pump power $P$ for the GaAs disk shown in Fig. 1. The lasing threshold $P \simeq 220 \mu$W. Inset: the emission spectrum at $P$ = 300 $\mu$W showing the lasing peak. (b) Spectral width $\Delta \lambda$ of the lasing peak as a function of pump power $P$. The straight lines are fitted curves.

\pagebreak
\begin{figure}[htbp]
\includegraphics[width=0.365\textwidth]{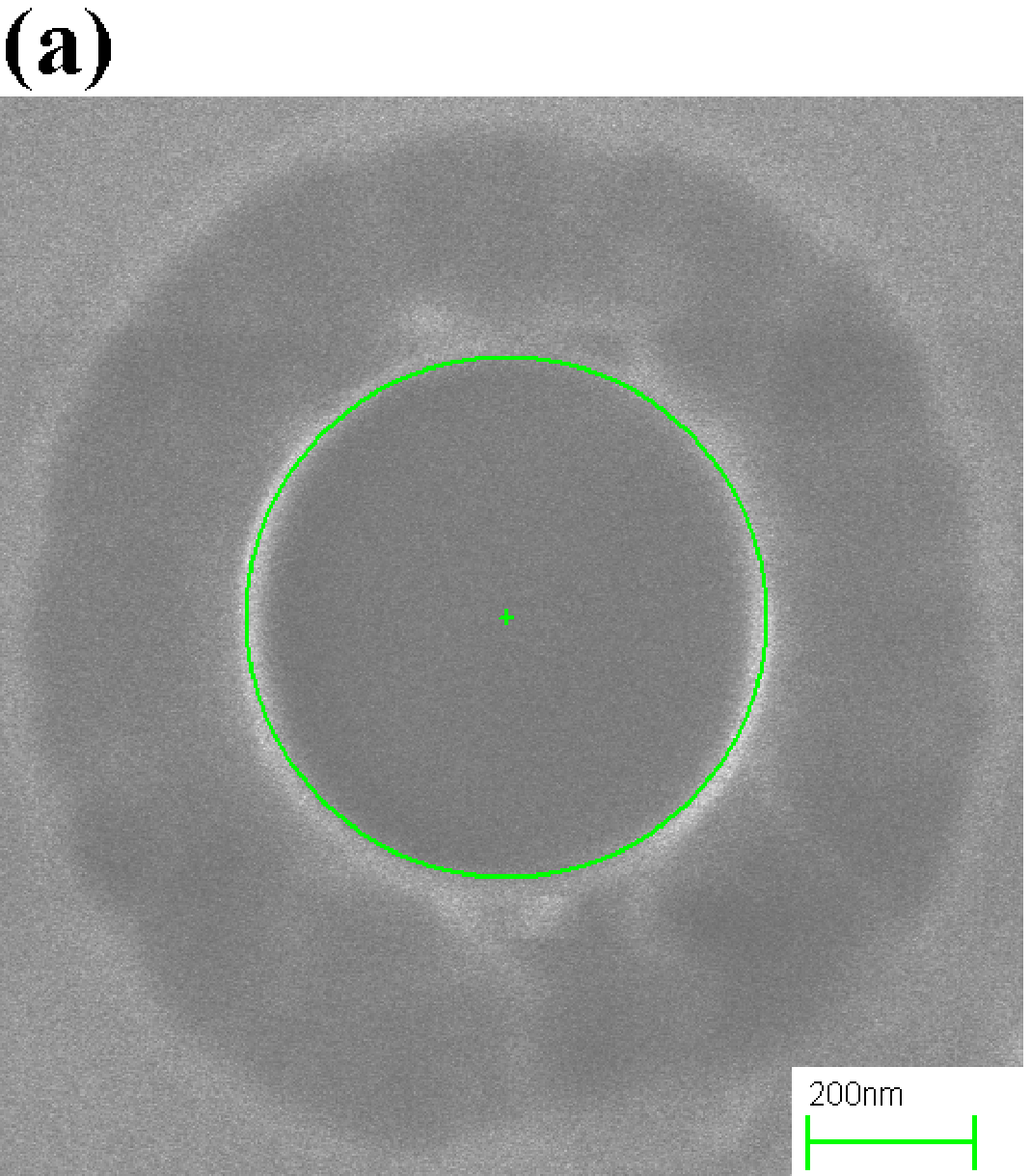}
\includegraphics[width=0.46\textwidth]{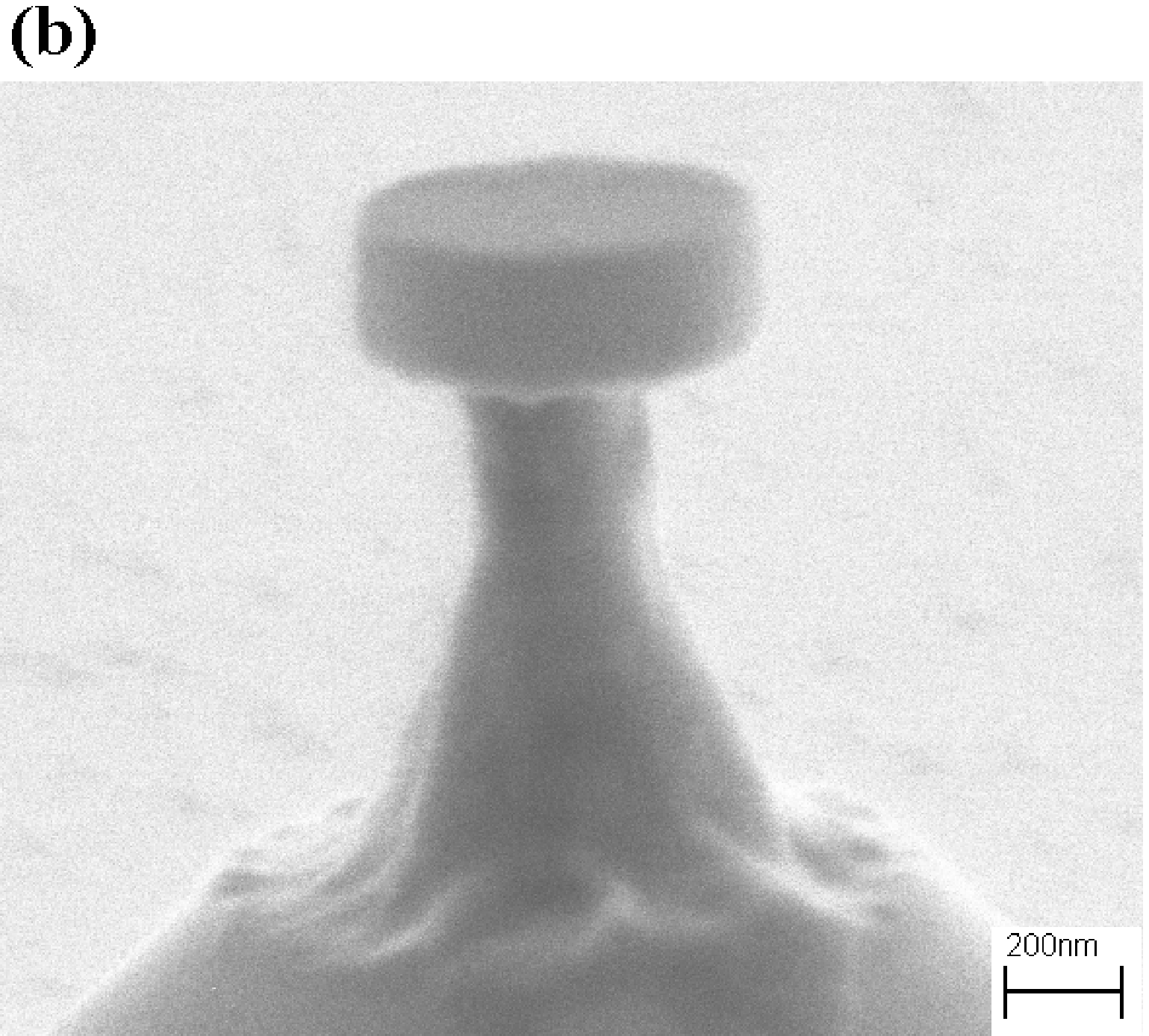}
\caption{ Song et al.}
\end{figure}

\pagebreak

\begin{figure}[htbp]
\includegraphics[width=0.4\textwidth]{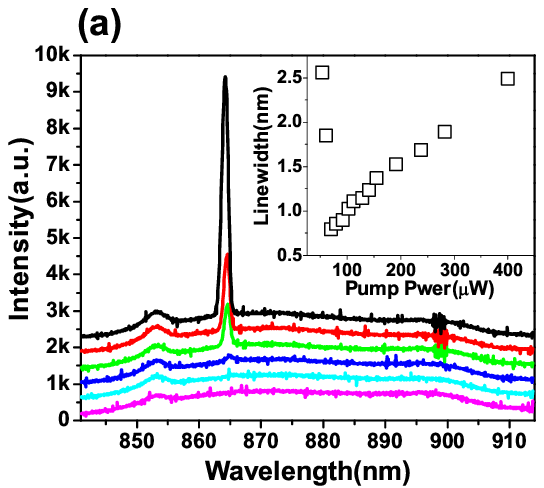}
\includegraphics[width=0.4\textwidth]{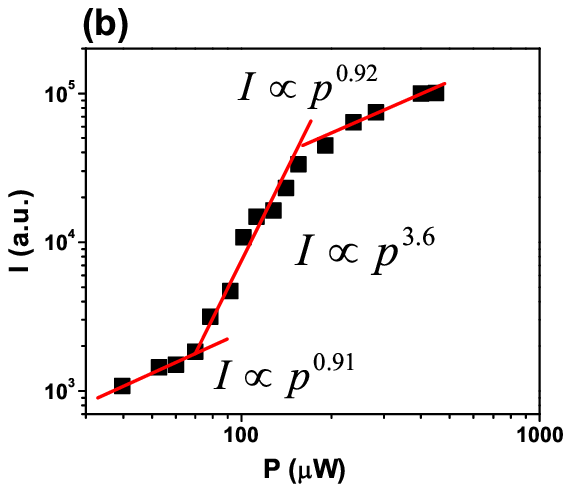}
\caption{ Song et al.} 
\end{figure}

\pagebreak

\begin{figure}[htbp]
\includegraphics[width=0.09\textwidth]{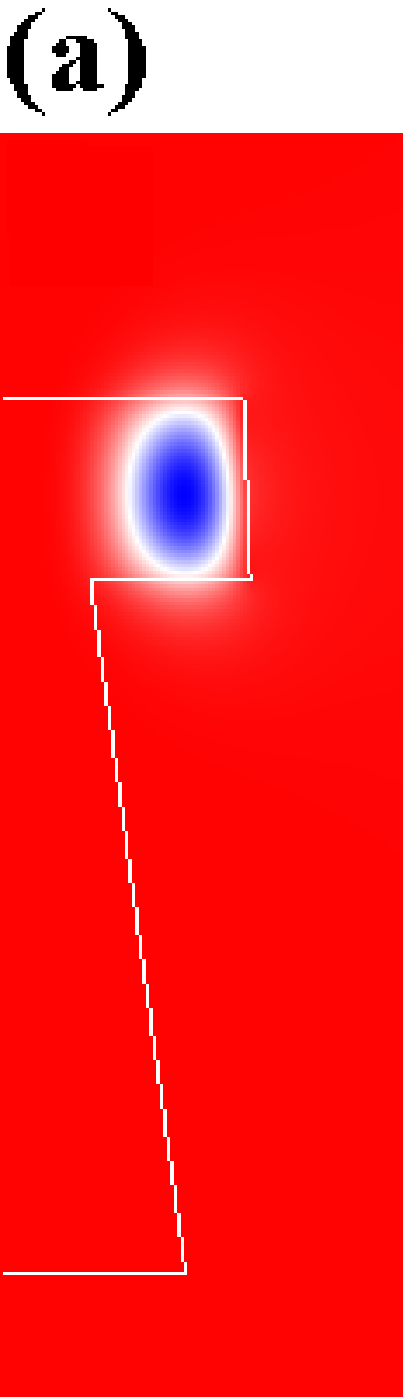}
\includegraphics[width=0.42\textwidth]{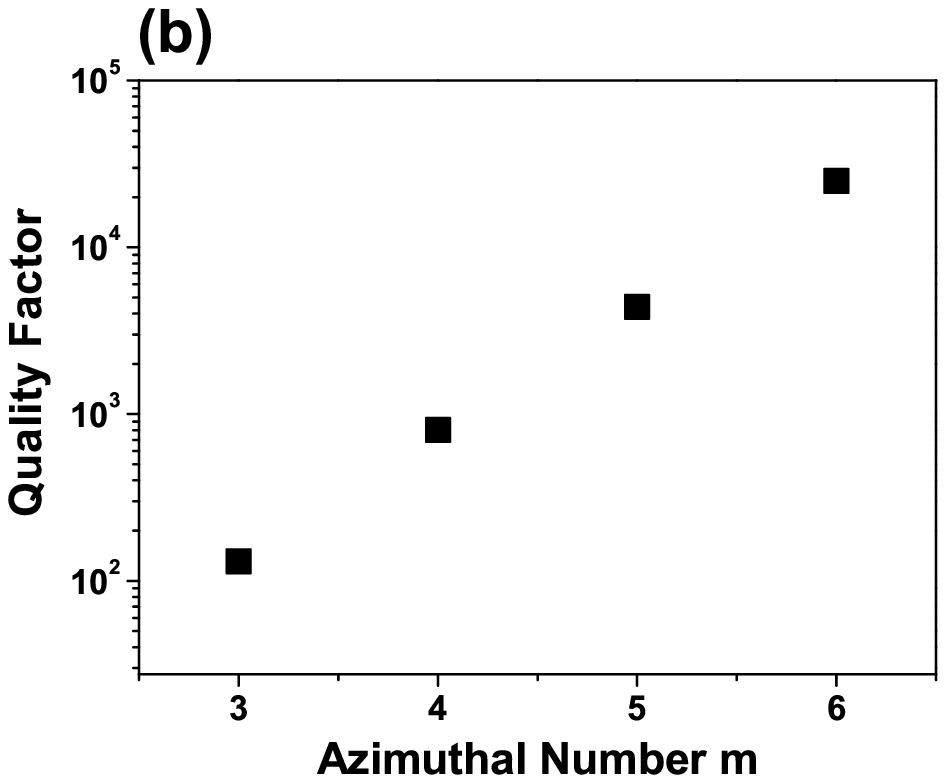}
\caption{ Song et al.  }
\end{figure}

\pagebreak

\begin{figure}[htbp]
\includegraphics[width=0.4\textwidth]{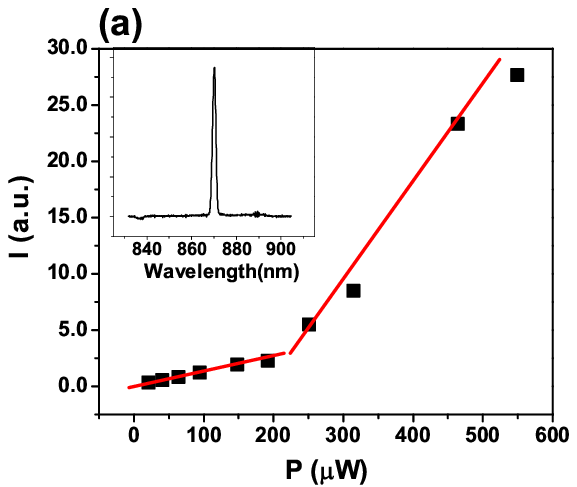}
\includegraphics[width=0.4\textwidth]{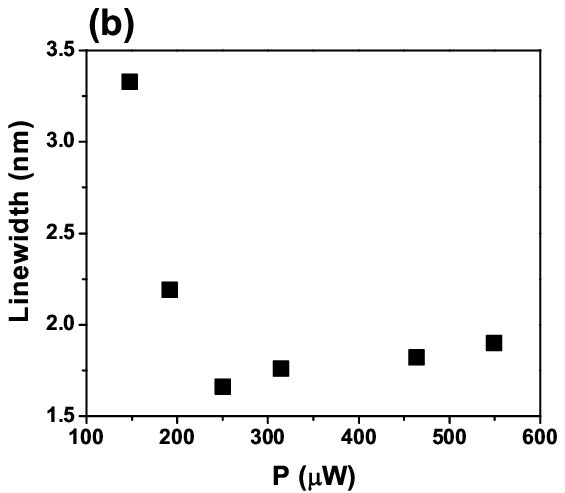}
\caption{ Song et al.  }
\end{figure}

\end{document}